\documentclass[preprint]{aastex6}
\turnoffedit
\usepackage{textgreek}
\def\wind{\textit{Wind}}
\def\stereo{\textit{STEREO}}
\def\sta{\stereo\textit{ A}}
\def\stb{\stereo\textit{ B}}
\def\stab{\sta{ and \textit{B}}}
\def\mes{\textit{MESSENGER}}
\def\sdo{\textit{SDO}}
\def\goes{\textit{GOES}}
\def\insitu{\textit{in situ}}
\def\fixb{F\textbeta}
\def\thecme{2012 July 12 CME}
\def\kmps{km s$^{-1}$}
\def\pcc{cm$^{-3}$}

\slugcomment{\textbf{Accepted for publication in The Astrophysical Journal}}
\shorttitle{Characteristics of the 2012 July 12 CME}
\shortauthors{Huidong Hu, Ying D. Liu, Rui Wang, et al.}

\begin{document}

\title{Sun-to-Earth Characteristics of the 2012 July 12 Coronal Mass Ejection
and Associated Geo-effectiveness}

\author{Huidong Hu\altaffilmark{1,2}, Ying D. Liu\altaffilmark{1,2},
	Rui Wang\altaffilmark{1}, Christian M\"ostl\altaffilmark{3,4},
	and Zhongwei Yang\altaffilmark{1}}
\altaffiltext{1}{State Key Laboratory of Space Weather,
	National Space Science Center,
	Chinese Academy of Sciences, Beijing 100190, China;\\
	\href{mailto:liuxying@spaceweather.ac.cn}{liuxying@spaceweather.ac.cn}}
\altaffiltext{2}{University of Chinese Academy of Sciences,
	No.19A Yuquan Road, Beijing 100049, China}
\altaffiltext{3}{Space Research Institute,
	Austrian Academy of Sciences, A-8042 Graz, Austria}
\altaffiltext{4}{IGAM-Kanzelh\"{o}he Observatory,
	Institute of Physics, University of Graz, A-8010 Graz, Austria}

\begin{abstract}
We analyze multi-spacecraft observations associated with the 2012 July 12
Coronal Mass Ejection (CME),
covering the source region on the Sun from \sdo,
stereoscopic imaging observations from \stereo,
magnetic field characteristics at \mes,
and type II radio burst and \insitu{} measurements from \wind.
A triangulation method based on \stereo{} stereoscopic observations
is employed to determine the kinematics of the CME,
and the outcome is compared with the result
derived from the type II radio burst
with a solar wind electron density model.
A Grad-Shafranov technique is applied to \wind{} \insitu{} data
to reconstruct the flux-rope structure
and compare it with the observation of the solar source region,
which helps understand the geo-effectiveness
associated with the CME structure.
Conclusions are as follows:
(1) the CME undergoes an impulsive acceleration,
a rapid deceleration before reaching \mes,
and then a gradual deceleration out to 1 AU,
which should be noticed in CME kinematics models;
(2) the type II radio burst was probably
produced from a high-density interaction region
between the CME-driven shock and a nearby streamer
or from the shock flank with lower heights,
which implies uncertainties in the determination
of CME kinematics using solely type II radio bursts;
(3) the flux-rope orientation and chirality deduced
from {\insitu} reconstruction at \wind{} agree with
those obtained from solar source observations;
(4) the prolonged southward magnetic field near the Earth
is mainly from the axial component
of the largely southward inclined flux rope,
which indicates the importance
of predicting both the flux-rope orientation
and magnetic field components in geomagnetic activity forecasting.
\end{abstract}


\section{Introduction} \label{intro}
Coronal Mass Ejections (CMEs) are massive expulsions of
magnetized plasma from the solar atmosphere.
They are called interplanetary CMEs (ICMEs)
after traveling into interplanetary space,
which are a significant class of triggers of geo-effectiveness.
Understanding CME propagation, associated radio bursts,
and plasma and magnetic field characteristics in the inner heliosphere
is of critical importance for space weather forecasting.
Combination of comprehensive remote-sensing and
\insitu{} observations is key to these investigations.

Previous studies of CME interplanetary propagation
categorize CMEs into fast and slow ones
by comparing their speed with the average solar wind speed.
Slow CMEs experience an acceleration
while fast ones decelerate when interacting with the ambient solar wind
\citep{sww1999,gll2000}.
Combining coronagraph images with \insitu{} measurements,
\citet{llr1999,gly2001} obtain
empirical models describing propagation of CMEs out to 1 AU.
\citet{gly2001}
find that a fast CME undergoes a deceleration out to 0.76 AU
before moving at a constant speed.
These studies are based on coronagraph images
with a field of view (FOV) only out to 30 {R$_\sun$}
and lack direct measurements between the Sun and Earth.
The \textit{Solar Terrestrial Relations Observatory}
\citep[\stereo;][]{kkd_stereo2008} with wide-angle imaging sensors
is capable of tracking CMEs from the Sun out to the Earth and even beyond.
\stereo{} consists of two spacecraft, one leading the Earth (\sta)
and the other trailing behind (\stb),
which separate by approximately 44 to 45 degrees from each other every year.
An identical imaging suite,
the Sun-Earth Connection Coronal and Heliospheric Investigation
\citep[SECCHI;][]{hmv_secchi2008},
is aboard each spacecraft,
which consists of an extreme ultraviolet imager (EUVI),
two coronagraphs (COR1 and COR2)
and two heliospheric imagers (HI1 and HI2).
A radio and plasma wave investigation instrument
\citep[SWAVES;][]{bgk_swaves2008} is also mounted,
which can detect type II radio bursts
associated with CME-driven shocks.
Based on coordinated \stereo{} stereoscopic images,
a geometric triangulation technique
has been developed to track CMEs with no free parameters
\citep{ldl2010,ltl2010}.
With the triangulation method,
it is revealed that fast CMEs impulsively accelerate
until even after the X-ray flare maximum,
and then rapidly decelerate to a nearly constant speed
or gradual deceleration phase
\citep{lll2013},
while slow CMEs experience a slow acceleration phase
and then travel with a roughly constant speed
around the average solar wind level
\citep{lhl2016}.
A CME could also propagate in a non-radial direction
\citep[e.g.,][]{wsw2004SoPh,gmx2009JGRA,mrf2015natco,lpo2015,wld2015apj},
interact with other CMEs
\citep[e.g.,][]{lvr2009AnGeo,gyk2001ApJ,llm2012,llk2014NatCo}
or co-rotating interaction regions
\citep[e.g.,][]{rkf1998,rls2010ApJ,lhl2016},
or rotate in interplanetary space
\citep[e.g.,][]{thv2006ApJ,ltl2010,vcn2011ApJ},
which increases difficulty
to predict the CME arrival characteristics at the Earth.

The \textit{MErcury Surface, Space ENvironment, GEochemistry, and Ranging}
\citep[\mes;][]{smg_mes2001} spacecraft
provides a great opportunity to study CMEs
inside the Earth orbit.
\citet{mfk2012} investigate
the shocks, flux ropes and interactions between ejecta
for a series of CMEs from 2010 July 30 to August 1
with multi-point \insitu{} data from \mes{},
\textit{Venus Express}, \wind{} and \stereo.
Radial evolution of a magnetic cloud (MC) is studied
based on the data from \mes{} and \stb{}
when the two spacecraft were nearly radially aligned in 2011 November
\citep{gfr2015}.
Using data from the \mes{} magnetometer,
\citet{wlp2015}
compile 61 ICMEs at Mercury between 2011 and 2014,
while
\citet{gf2016solphys} identify 119 ICMEs combining the data
from \mes{} and \textit{Venus Express}.

Interplanetary type II radio bursts emit at the fundamental and/or harmonic
of the plasma frequency,
which can be applied to determine the radial distance
and predict the time of arrival (ToA) of CME-driven shocks at 1 AU
using a proper electron density model of the solar wind
\citep{rkb2007,llm2008,lll2013,cis2015}.
Using the electron density model of
\citet{ldb1998},
\citet{lll2013}
derive the radial distances of CME-driven shocks
from the drift rates of type II bursts,
which compare well with those acquired from
the geometric triangulation technique
based on \stereo{} stereoscopic white-light observations.
Besides from the nose of a CME-driven shock,
type II radio bursts can also be produced from the flank
\citep{ca2002,clm2007},
or a preexisting high-density solar wind structure interacting with the shock
\citep[e.g.,][]{rkf1998,rvc2003ApJ,clm2007,fck2012ApJ}.
These complicate the estimate of the radial distance of a shock
based on an electron density model.

Prolonged and enhanced southward magnetic fields
associated with ICMEs are important triggers of geomagnetic storms,
which depend on the flux-rope orientation
and the axial and azimuthal magnetic field components
\citep{ywg2001,lhw2015}.
A statistical study finds that
the tilt angles of ICME flux ropes
deduced from \insitu{} force-free flux-rope models
are close to the tilt angles
of magnetic polarity inversion lines (PILs)
in the corresponding solar source regions
\citep{may2015}.
A Grad-Shafranov (GS) reconstruction technique
is capable of estimating the flux-rope orientation,
and axial and azimuthal magnetic field components
from \insitu{} measurements
\citep{hs1999,hs2002},
which has been validated by well separated multi-spacecraft measurements
\citep{llh2008,mfm2009}.
With the GS reconstruction technique,
\citet{lhw2015}
find that the flux-rope axial component
is a major contributor to the southward magnetic field
for the 2015 June 22 geomagnetic storm
while the azimuthal component plays a major role
in the 2015 March 17 strong geomagnetic storm,
which indicates that the flux-rope structure plays an important role
in the generation and intensity of geomagnetic activity.

On 2012 July 12, a major CME erupted in NOAA AR 11520
associated with an X1.4 class flare that peaked at about 16:49 UT.
The active region is the same
that caused the 2012 July 23 extreme solar storm
\citep{llk2014NatCo}.
A strong geomagnetic storm was triggered
and reached a minimum D$_\mathrm{st}$ index of $-$127 nT on July 15.
The magnetic field configuration and evolution in the solar source region
have been investigated with remote-sensing observations
and/or three-dimensional magnetohydrodynamics (3D MHD) simulations
\citep{cdz2014ApJ,dja2014ApJ,scz2015ApJ,wlw2016}.
The CME kinematic evolution has been studied
with a 3D MHD model and a semi-empirical drag model by
\citet{ssz2014JGRA} and
\citet{hz2014ApJ}, respectively.
This event is also included as an illustration in statistical analyses of
\citet{mah2014ApJ} and \citet{wlp2015}.

In this paper we present a comprehensive remote-sensing
and \insitu{} analysis of the \thecme.
Despite a number of previous studies of this event,
our work is unique for reasons given below:
(1) a triangulation method based on stereoscopic wide-angle imaging observations
is employed to determine the CME kinematics in this study,
while the previous studies are limited to single-spacecraft analysis;
(2) the stereoscopic imaging results
are compared with multi-point \insitu{} measurements
at Mercury and the Earth;
(3) the frequency drift rate of the type II radio burst
associated with the event is analyzed in this work,
from which the results are compared with the triangulation analysis;
(4) we apply the GS technique to the near-Earth \insitu{} data
to reconstruct the flux-rope structure,
and the result is compared with the solar source observations
to understand the association of the flux-rope properties
with the generation of the geomagnetic storm.
As far as we know,
the comparison between stereoscopic wide-angle imaging observations
and multi-point \insitu{} measurements,
which can give crucial information on CME kinematics and structure,
is still lacking.
The comparison of wide-angle imaging observations
with long-duration interplanetary type II radio bursts
can also yield important knowledge of CME kinematics
as well as source regions of the type II bursts,
which has not been sufficiently studied.
The examination of the CME magnetic structure and
its comparison with the analysis of the solar source region
are key to understanding
how the CME structure/solar source signature
is connected with geomagnetic activity.
We describe the solar source signatures in Section \ref{source}
and propagation characteristics
in interplanetary space in Section \ref{prop}.
Section \ref{earth} examines the properties of the flux rope
and the associated geo-effectiveness.
These results are summarized and discussed in Section \ref{discuss}.
This work illustrates an end-to-end study of the space weather chain
from the Sun to Earth,
highlighting the importance of multi-spacecraft
remote-sensing and \insitu{} observations
in understanding the Sun-to-Earth characteristics
and geo-effectiveness of CMEs.

\section{Source Region Signatures} \label{source}
\begin{figure}
\epsscale{.80}
\plotone{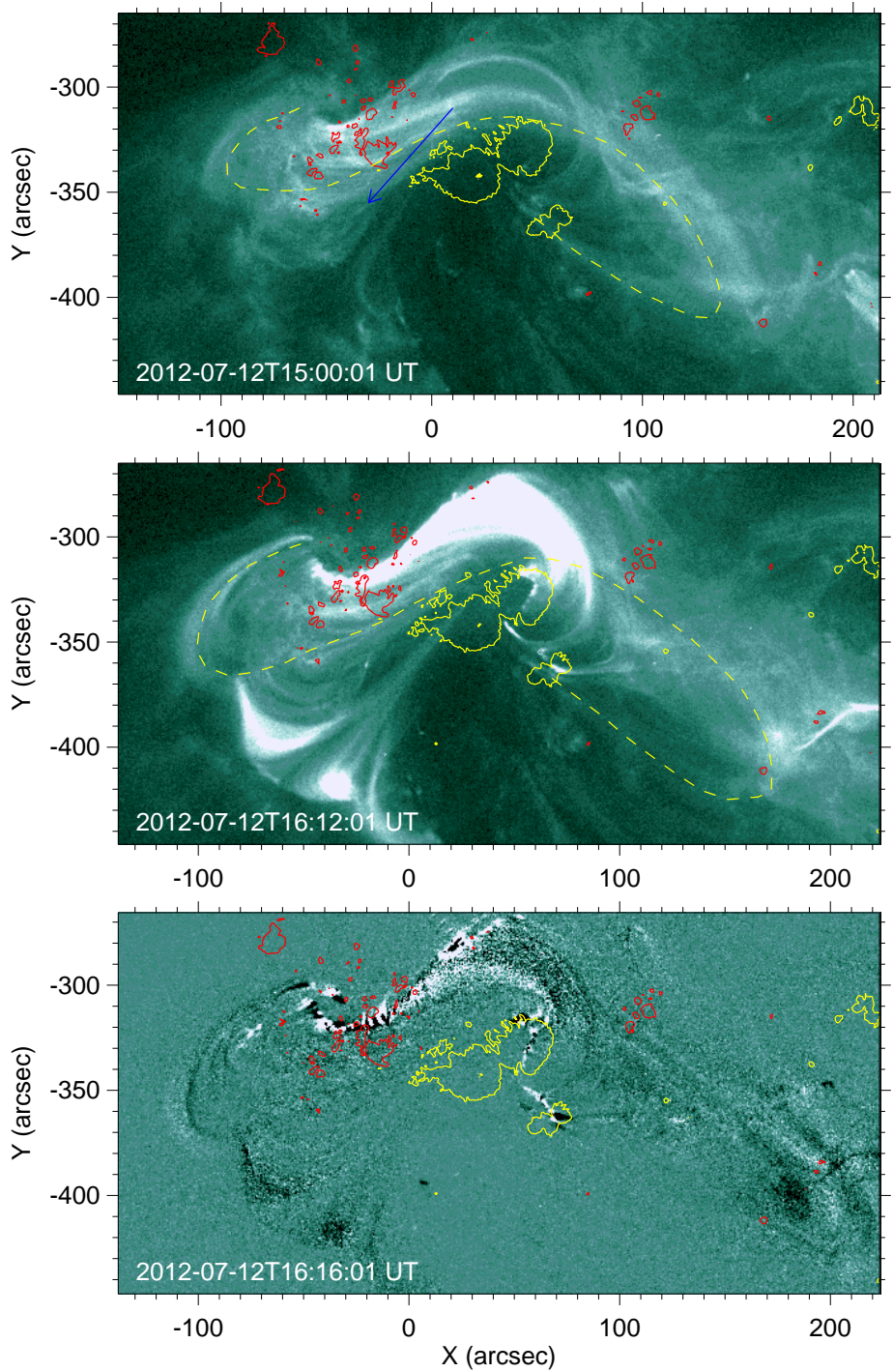}
\caption{\label{fig1}\sdo/AIA 94 {\AA} EUV images
of NOAA AR 11520 before the eruption.
The bottom panel shows an EUV running difference image
with time delay of 1 minute.
The dashed yellow curves in the top and middle panels
indicate the synoptic configuration
of the flux rope which produced the CME.
The bright pattern in the middle panel
is the low-lying flux rope that did not erupt in the event
\citep{cdz2014ApJ}.
The yellow and red contours represent the areas of the magnetic field
along the line of sight larger than 1000 Gauss
for positive and negative polarities, respectively.
The blue arrow in the top panel
roughly marks the position and direction
of the polarity inversion line near the central meridian.}
\end{figure}

The CME was launched from NOAA AR 11520 (source location S15{\degr}W01{\degr}),
the same active region that spawned the 2012 July 23 extreme solar storm
\citep{llk2014NatCo}.
It was associated with an X1.4 class flare that peaked
around 16:49 UT on 2012 July 12.
\sdo/AIA
\citep{lta_aia2012} 94 {\AA} EUV images at three moments
before the eruption are displayed in Figure \ref{fig1}.
The hot channel marked by the dashed yellow curve
is identified to be the erupted flux rope forming the CME
\citep{cdz2014ApJ,dja2014ApJ,scz2015ApJ}.
The running difference image in the bottom panel
clearly indicates the expansion of the flux rope
as reported by \citet{dja2014ApJ}.
The PIL near the central meridian is approximately
from the northwest to the southeast
according to the contours from a {\sdo}/HMI
\citep{ssb_hmi2012} line-of-sight magnetogram.
The axial magnetic fields of the flux rope
indicated by the hot channel,
which start from the positive polarity,
are generally pointing to the southeast near the central meridian.
The overlying azimuthal components should be
from the right side (positive polarity)
to the left side (negative polarity)
in the view from the Earth.
This magnetic configuration thus forms a right-handed flux rope,
which is also consistent with a statistical research
that the southern hemisphere tends to produce positive magnetic helicity
\citep{pcm1995ApJ}.
We will compare the flux-rope chirality and orientation
with the \insitu{} reconstruction results at \wind.

\section{Propagation in Interplanetary Space} \label{prop}
We derive the kinematics of the {\thecme} using
a geometric triangulation technique proposed by
\citet{ldl2010,ltl2010}
based on stereoscopic imaging observations from {\stereo}.
The triangulation method adopts
two geometric assumptions for the CME leading edge,
called fixed \textbeta{} (\fixb) and harmonic mean (HM)
approximations, respectively.
The \fixb{} approximation assumes
that a CME is a relatively compact structure
simultaneously observed by the two spacecraft
when it is moving away from the Sun
\citep{sww1999,kw2007,ltl2010}.
The HM approximation assumes that the leading edge of a CME
is a sphere which is attached to the Sun
and tangent to the lines of sight of the two spacecraft
\citep{lvr2009AnGeo,ltl2010}.
A self-similar expansion model
expressing the same triangulation concept is developed by
\citet{dpt2013ApJ},
for which the \fixb{} and HM approximations are limiting cases.
The triangulation concept has proved to be useful
for acquiring CME Sun-to-Earth kinematics
and connecting remote sensing observations with \insitu{} signatures
\citep[e.g.,][]{ldl2010,ltl2010,lll2013,lhr2010,
mtr2010,hdm2012,tvr2012,dpt2013ApJ}.
Detailed descriptions and discussions of the triangulation technique
can be found in
\citet{ltl2010,lll2013,lhl2016}.

\begin{figure}
\epsscale{.80}
\plotone{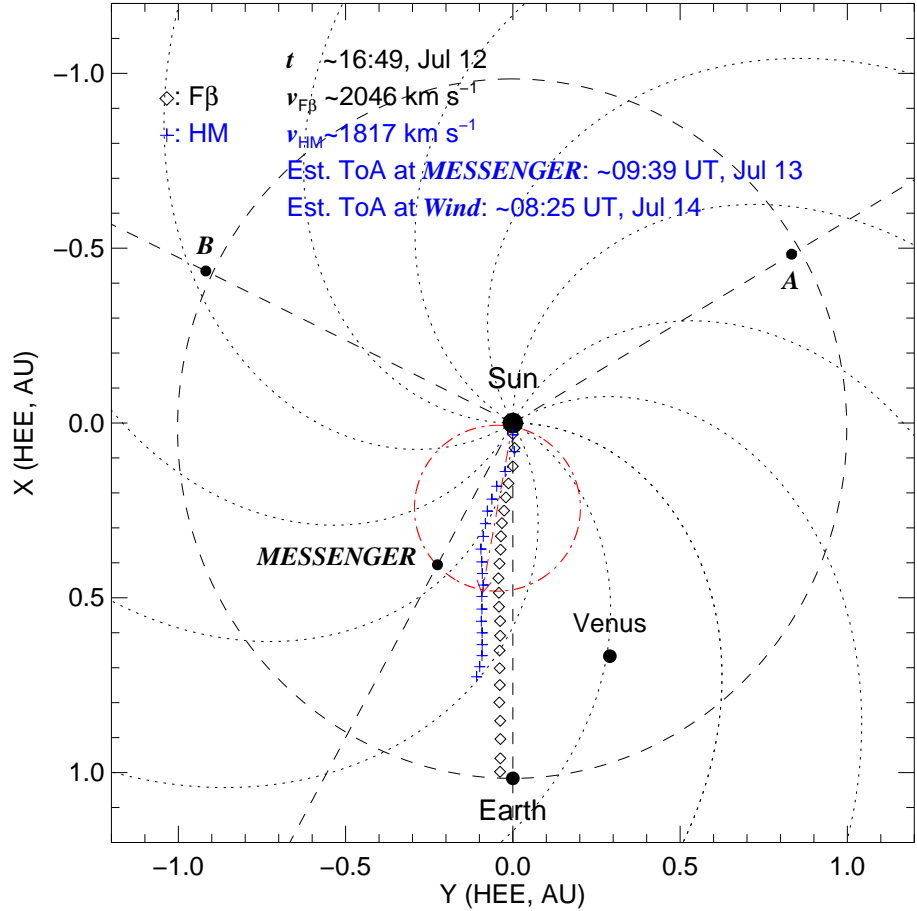}
\caption{\label{fig2}Positions of the spacecraft and planets
in the ecliptic plane on 2012 July 13.
The trajectories of the {\thecme} are obtained from triangulations
with the \fixb{} (black diamond)
and HM (blue cross) approximations, respectively.
The red circle represents the size
of the assumed spherical CME leading edge when the shock arrives at \mes{},
and the red arrow indicates the direction then.
The black circle marks the orbit of the Earth,
and the gray dotted curves show Parker spiral magnetic fields
created with a solar wind speed of 450 \kmps.
The estimated CME launch time on the Sun
and derived peak speeds are given.
The estimated arrival times of the shock
at \mes{} and \wind{} by the HM approximation are printed in blue.}
\end{figure}

Figure \ref{fig2} shows the trajectories of the \thecme{}
obtained from the {\fixb} and HM triangulations
as well as the positions of the planets and spacecraft
in the ecliptic plane on 2012 July 13.
\sta{} was $\sim$120{\degr} ahead of the Earth
and \textit{B} was $\sim$115{\degr} behind.
Venus was $\sim$0.73 AU from the Sun and $\sim$23.5{\degr}
west of the Earth,
while {\mes} (Mercury) was $\sim$0.47 AU from the Sun
and $\sim$30.6{\degr} east of the Earth.
This is a halo CME that impacted Mercury ({\mes}), the Earth (\wind)
and possibly Venus.
The maximum speed of the CME near the Sun was
$\sim$2046 and $\sim$1817 {\kmps} estimated
by the {\fixb} and HM triangulations, respectively.

\begin{figure}
\epsscale{.80}
\plotone{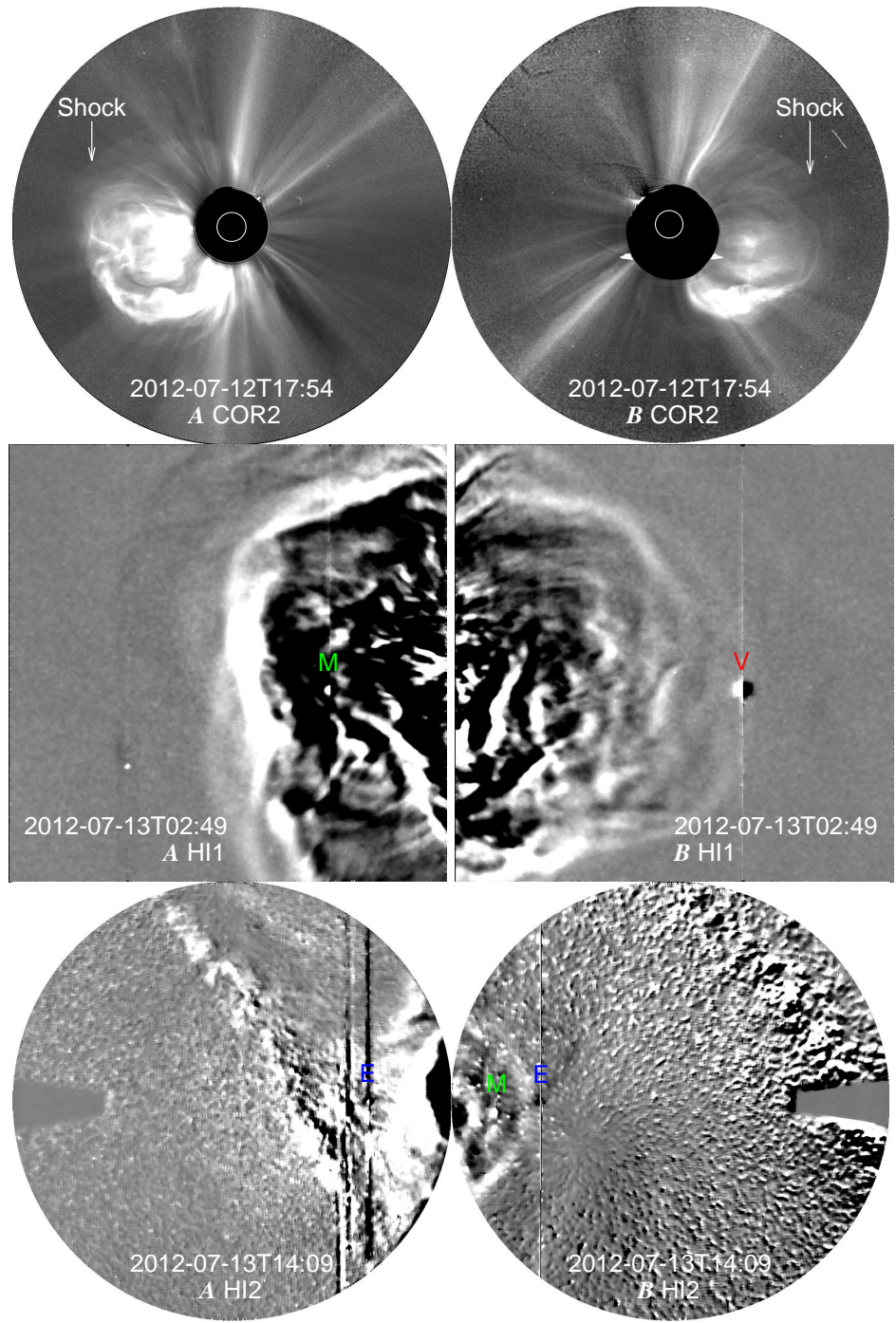}
\caption{\label{fig3}Evolution of the {\thecme}
viewed simultaneously from {\sta} (left) and \textit{B} (right).
From top to bottom, the panels show the images of COR2,
running difference images of HI1 and HI2, respectively.
The shock driven by the CME is visible in the COR2 images.
The positions of Mercury (M), Venus (V) and the Earth (E)
in the fields of view are marked in corresponding HI images.}
\end{figure}

\begin{figure}
\epsscale{.80}
\plotone{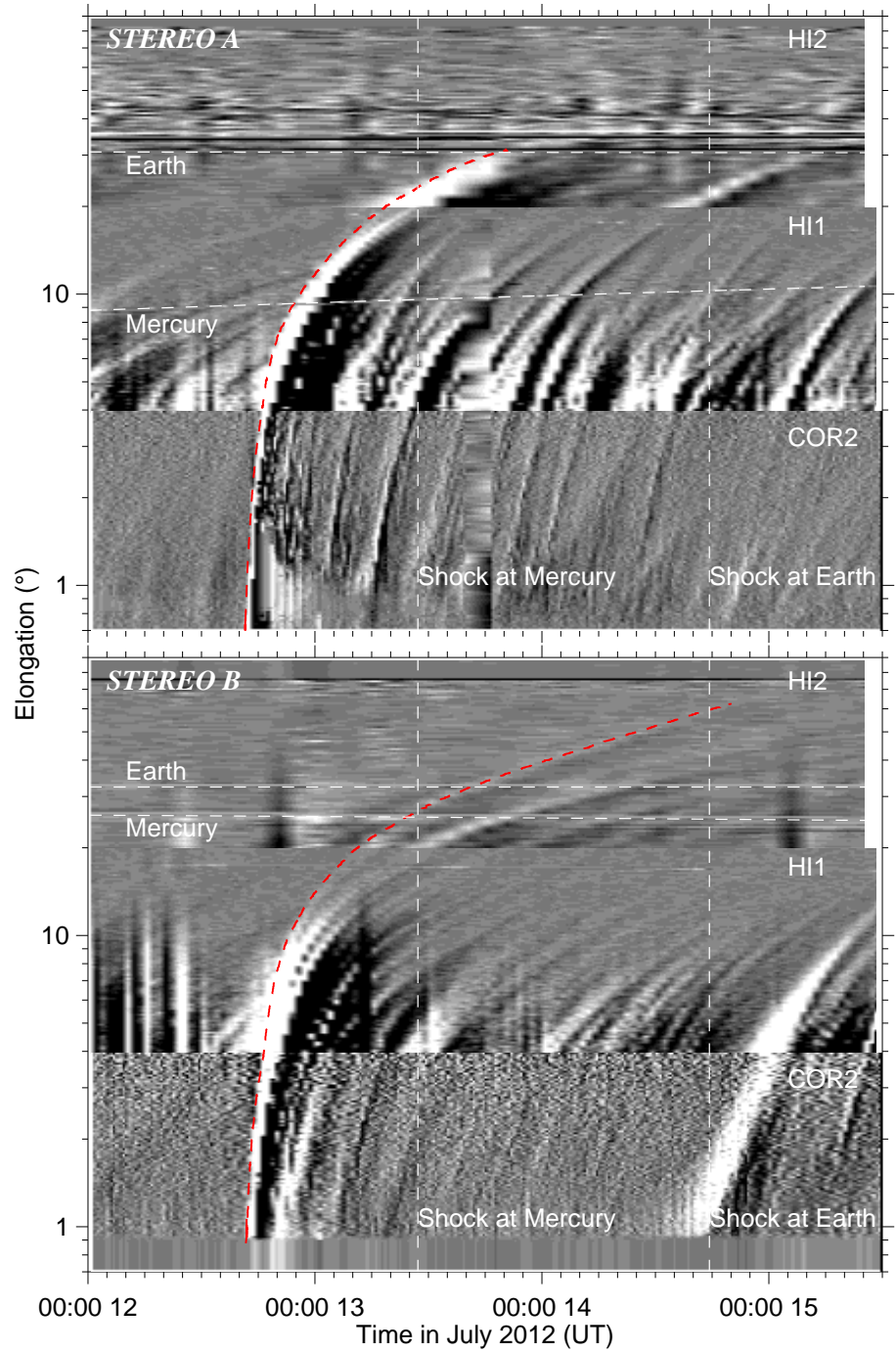}
\caption{\label{fig4}Time-elongation maps constructed
from running difference images along the ecliptic plane.
The red dashed curve indicates the CME track,
along which the elongation angles are extracted.
The vertical dashed lines mark the observed arrival times of the shock
at Mercury and the Earth.
The horizontal dashed lines denote the elongation angles
of the Earth and Mercury.}
\end{figure}

The {\stereo} spacecraft tracked
the whole propagation of the CME from the Sun to Earth
with white-light observations from the SECCHI instruments.
Figure \ref{fig3} depicts the evolution of the CME
from COR2, HI1 and HI2 of {\stab}.
COR2 has a 0.7\degr--4{\degr} FOV around the Sun.
HI1 has a 20{\degr} square FOV centered at 14{\degr} elongation
from the center of the Sun,
while HI2 has a 70{\degr} FOV centered at 53.7{\degr}.
The shock ahead of the CME ejecta is visible in the COR2 images,
which was interacting with nearby streamers.
By stacking the running difference intensities
of COR2, HI1, and HI2 within a slit along the ecliptic plane,
we obtain two time-elongation maps
\citep[J-maps, e.g.,][]{shp2008ApJ,dhr2009GeoRL,ldl2010}
as shown in Figure \ref{fig4}.
The time when the CME track intersected with the Earth elongation line
in the view of \stb{} is a little earlier than that for \sta,
which indicates that the CME was propagating in a direction slightly
to the east of the Sun-Earth line
(see Figure \ref{fig2}).
Due to the contamination by the Milky Way galaxy
in \sta{ HI2} as displayed in Figure \ref{fig3},
the track duration from \sta{} is shorter than that from \stb.

\begin{figure}
\epsscale{.80}
\plotone{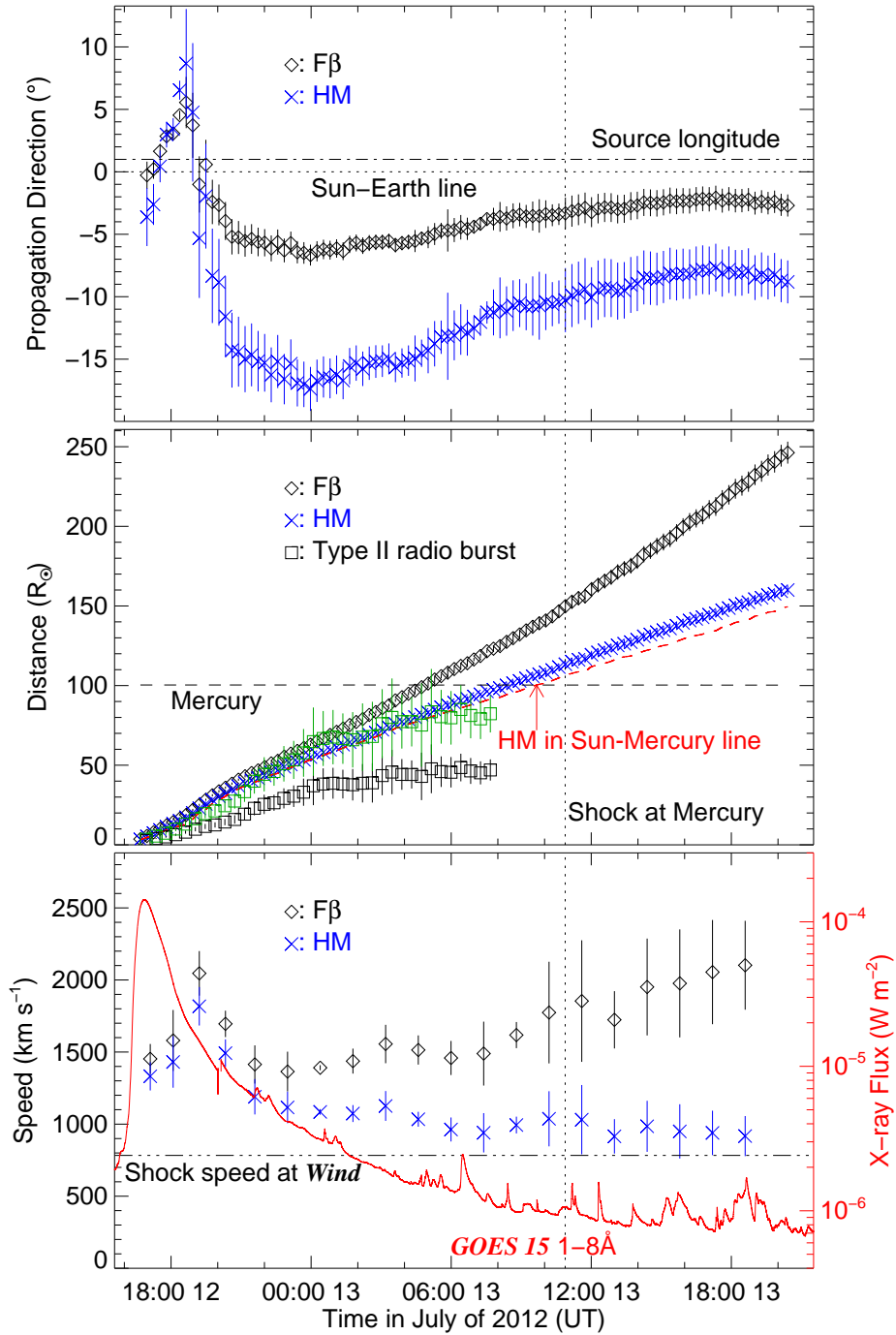}
\caption{\label{fig5}Propagation direction, radial distance
and speed profiles of the leading edge of the {\thecme} derived
from triangulations with the {\fixb} (black diamond)
and HM (blue cross) approximations.
The Sun-Earth line and the longitude of the CME source location
on the Sun are indicated by the horizontal lines in the top panel.
In the middle panel,
the black squares denote the distances derived
from the associated type II radio burst
using the Leblanc model
with an electron density of 6.5 {\pcc} at 1 AU,
while the green squares are corresponding distances
obtained with an electron density of 20 {\pcc} at 1 AU.
The red dashed curve indicates the HM triangulated distance
of the CME leading edge along the Sun-Mercury line,
and the horizontal line in the middle panel marks the distance of Mercury.
The speeds are calculated from adjacent distances
using a numerical differentiation with three-point Lagrangian interpolation
and are then binned to reduce the scatter.
The horizontal line in the bottom panel
indicates the shock speed measured at \wind,
and the red curve is the {\goes} X-ray flux scaled by the red axis on the right.
The vertical dotted line marks the arrival time of the shock at \mes.}
\end{figure}

The CME kinematics in the ecliptic plane derived
from the triangulation method is plotted in Figure \ref{fig5}.
The average CME directions during the whole propagation process
obtained from the two approximations are $\sim$4{\degr} and $\sim$12{\degr}
to the east of the Sun-Earth line, respectively,
which are consistent with the results obtained from the single-spacecraft
self-similar expansion (SSEF) and harmonic mean (HMF) fittings
\citep{mah2014ApJ}.
The angle derived from the HM triangulation
is roughly twice the one from the {\fixb} triangulation and more variational,
which is also noticed in previous studies
\citep{lhr2010,lll2013,lhl2016}.
The distance derived from the {\fixb} approximation
becomes larger than that from the HM approximation around 50 R$_\sun$
and the speed from the {\fixb} approximation
starts to show an unphysical acceleration around 100 R$_\sun$,
which are also noted in previous studies
\citep{lvr2009AnGeo,whp2009,lll2013}
and due to the non-optimal observation geometry of the two spacecraft
(i.e., observing from behind the Sun)
in combination with the \fixb{} restriction
\citep{lll2013,lhl2016}.
The speed trends from the {\fixb} and HM approximations
below $\sim$50 R$_\sun$ are similar,
except that the peak speed from {\fixb} is $\sim$200 {\kmps} larger.
Both the two speed profiles show that the CME accelerates
out to $\sim$20 R$_\sun$ even after the X-ray flux peak time
and then rapidly drops to $\sim$1400 ({\fixb})
and $\sim$1200 (HM) {\kmps} at $\sim$50 R$_\sun$ in about 4 hours.
The speeds during the acceleration phase
are consistent with that
obtained from the Graduated Cylindrical Shell (GCS) model by
\citet{mah2014ApJ},
while the latter is an average from 2.5 to 15.6 R$_\sun$
which is still in the acceleration phase.
The {\fixb} speed after the acceleration phase
is consistent with the constant speed of 1486 {\kmps}
derived from the SSEF fitting by
\citet{mah2014ApJ},
while the HM counterpart is smaller.
Ignoring the unphysical acceleration from the {\fixb} approximation,
we can see that the CME undergoes a gradual deceleration phase
after the peak speed.
Both the speed profiles acquired from the {\fixb} and HM triangulations
are similar to those of the three fast CMEs in
\citet{lll2013}.

\begin{figure}
\epsscale{.80}
\plotone{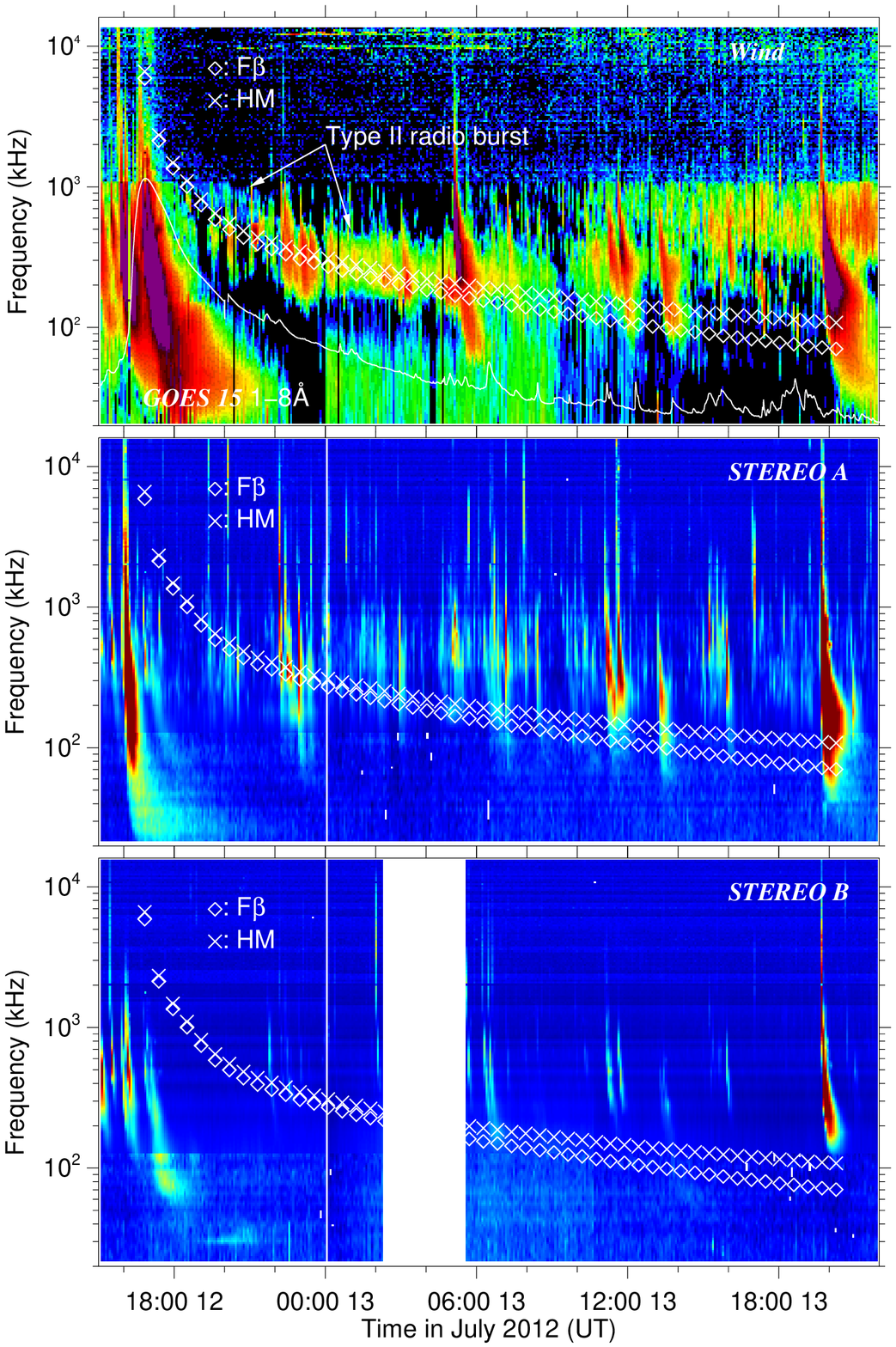}
\caption{\label{fig6}Radio dynamic spectra
associated with the \thecme{} from \wind, \stab.
The CME leading edge distances derived
from triangulation with the \fixb{} (diamond) and HM (cross) approximations
are converted to frequencies using the Leblanc model
with $n_e=20$ \pcc{} at 1 AU,
and then plotted over the dynamic spectra.
\goes{} X-ray flux is also overlapped in the {\wind} plot
scaled in arbitrary units.
The white area in the bottom panel indicates the data gap.}
\end{figure}

The radio dynamic spectra associated with the CME
from \wind{} and \stereo{} are shown in Figure \ref{fig6}.
A long-duration type II radio burst is observed only by \wind,
which starts from the upper boundary of 16 MHz
at $\sim$17:00 UT on July 12 after the X-ray flux peak
and drifts to $\sim$250 kHz at $\sim$9:00 UT on July 13.
The drift rate from $\sim$300 kHz
becomes more gradual around 2:00 UT on July 13.
There are several solar wind electron density models
\citep[e.g.,][]{fs1971SoPh,spm1977SoPh,bks1984SoPh}
that can interpret the frequency of type II radio as
the distance of emission source.
We choose a popular electron density model derived by
\citet[][referred to as the Leblanc model hereafter]{ldb1998}
that covers a range from about 1.8 R$_\sun$ to 1 AU
embracing the whole triangulation derived CME propagation distance.
Results of three CME events from this model
agree well with those from the same triangulation method
as employed in this work
\citep{lll2013}.
The CME leading edge distances obtained from the triangulation
with the \fixb{} and HM approximations
are converted to frequencies
using the Leblanc model
with an electron density $n_e=20$ \pcc{} at 1 AU,
and are then doubled to their harmonic frequencies
and plotted over the spectra.
The frequencies from both the \fixb{} and HM triangulations
underestimate the observed type II radio band
in the initial phase (before $\sim$22:00 UT on July 12) and
final phase (after $\sim$4:00 UT on July 13),
which suggests that the distances of the type II radio source regions
are smaller than those derived from the triangulation method
during the two time ranges.
Note that the electron density of 20 {\pcc} at 1 AU is already very high
compared with the average solar wind electron density at 1 AU,
and it will require an even higher electron density
if we assume the type II radio emission at the fundamental frequency.

Assuming that the emission is produced at the second harmonic frequency,
we derive the shock distance from the type II radio burst
using the Leblanc model.
The distances obtained from the electron density
of $n_e=6.5$ and $n_e=20$ \pcc{}
at 1 AU are plotted in the middle panel of Figure \ref{fig5}.
The uncertainties are obtained from the width of the type II radio band
shown in Figure \ref{fig6}.
The distance corresponding to $n_e=6.5$ {\pcc} at 1 AU
is much smaller compared with the distance
derived from the HM triangulation.
This indicates that the electron density
inferred from the radio frequency is much higher
than that from the Leblanc model with $n_e=6.5$ {\pcc} at 1 AU.
The more gradual increase of the distance from $\sim$40 to $\sim$50 R$_\sun$
suggests that the frequency drift during that time period is slower.
In order to match the triangulation results,
a larger electron density $n_e=20$ {\pcc} is used,
from which the distance is roughly consistent with the triangulation results
between 22:00 UT on July 12 and 4:00 UT on July 13
but is still smaller than the triangulation results outside the time range.
These results imply that the type II radio burst
was probably produced from the shock flank with lower heights
or a high-density streamer interacting with the shock.
We see more than one streamers interacting with the shock in Figure \ref{fig3},
but we cannot determine which interaction produced the type II radio burst.

Figure 7 shows the solar wind magnetic field measurements
by the \mes{ Magnetometer}
\citep{aal_mag2007}.
\mes{} was orbiting Mercury approximately in the Y-Z plane of
the Mercury Solar Orbital frame with a period of 8 hours,
the periapsis of 2795 km and the apoapsis of 12677 km
(from the center of the planet).
Data below 4 Mercury radii (9760 km),
where the measurements are supposed to be dominated by the magnetosphere,
are excluded.
A shock with a sharp increase in the magnetic field strength
was observed at 10:53 UT on July 13 before the apoapsis.
The HM triangulation predicts an arrival time of the CME leading edge
at {\mes} around 09:39 UT on July 13 (see Figure \ref{fig5})
which is only about 1 hour earlier than observed at \mes.
We cannot estimate the flux-rope orientation and chirality
since the magnetic field was dominated by the expanded Mercury magnetosphere
for most of the time during the event.
Note that the rapid deceleration occurred
before the CME reached {\mes}
and thereafter the CME speed is roughly constant (see Figure \ref{fig5}).
The CME also plausibly impacted \textit{Venus Express},
since we see the putative end of the flux rope on July 15
through the magnetic field data from the spacecraft.
However, we cannot determine the arrival time of the shock
because there is a data gap during the CME arrival.

\section{Near-Earth Properties and Geo-effectiveness} \label{earth}
The \insitu{} measurements associated with the \thecme{}
at \wind{} are presented in Figure \ref{fig8},
showing that a shock passed \wind{} at 17:43 UT on July 14.
Due to the noisy backgrounds in \stb{ HI2} images
(see Figure \ref{fig3}),
the HM triangulation can track the CME
out to $\sim$150 R$_\sun$ (see Figure \ref{fig5}).
We use a linear fit of the distances
and assume a propagation direction of $-10${\degr} near 1 AU
to estimate the shock ToA at \wind.
The fit predicts a ToA of 08:25 UT on July 14
which is about 9 hours earlier than the observed shock arrival,
and a speed of $\sim$930 {\kmps}
which is larger than the shock speed
($\sim$770 {\kmps}) observed at \wind.
A flux-rope-like structure with an interval
from 7:38 UT on July 15 to 14:28 UT on July 16
is identified through the boundary sensitive GS reconstruction method
(see the text below)
in combination with the plasma and magnetic field parameters.
In the flux rope, the magnetic field strength decreases smoothly,
the R component increases from $-$20 nT to around zero,
the T component rotates from negative to positive,
and the N component keeps negative and increases from $-$18 to $-$9 nT.
There is possibly another flux rope
following the reconstructed one with a different magnetic configuration
extending to $\sim$04:30 UT on July 17 in the MC,
which, however, can not be reconstructed by the GS method.
Following the sudden commencement at the shock arrival,
a major geomagnetic storm occurred because of the southward magnetic field,
with the minimum D$_\mathrm{st}$ of $-$127 nT.
Both the
\citet{om2000JGR} and \citet{bmr1975JGR} D$_\mathrm{st}$ models
underestimate the D$_\mathrm{st}$ measurements.
This is possibly due to the low solar wind density
within the MC.
Previous studies reveal that a high density
can intensify the ring current
by feeding the plasma sheet of the magnetosphere
\citep{fjt2006JGRA,ltb2006JGRA}.
The geomagnetic storm would have been stronger
if the density inside the MC were higher.

We obtain a cross section of a flux-rope-like structure
as shown in Figure \ref{fig9} using the GS reconstruction
\citep{hs1999,hs2002},
which helps understand
how the CME structure contributes to the geomagnetic storm activity.
The magnetic fields are in a flux-rope frame
where $x$ is nearly along the spacecraft trajectory
and $z$ in the direction of the flux-rope axis.
The structure is right-handed as inferred
from the transverse fields along the spacecraft trajectory.
The elevation angle of the flux-rope axis is about $-$44{\degr}
and the azimuthal angle is about 232{\degr} in RTN coordinates,
as determined from the single-valued behavior
of the transverse pressure over the vector potential
\citep[][see below]{hs2002}.
The elevation angle is comparable to the GCS results reported by
\citet{mah2014ApJ} and \citet{hz2014ApJ}
based on coronagraph images.
The orientation and chirality of the reconstructed flux rope
are consistent with those derived from the solar source observations
displayed in Figure \ref{fig1}.
The maximum strength of the axial magnetic field is $\sim$30 nT,
about two times as large as the maximum value
of the azimuthal component (13 nT).
Therefore, the axial magnetic field component
on top of the largely southward orientation of the flux rope
is the main contributor
to the southward magnetic field triggering the geomagnetic storm.

The plot of the transverse pressure $P_\mathrm{t}$
versus the vector potential $A(x,0)$\textbf{\^{z}} in Figure \ref{fig10}
indicates the reliability of the GS reconstruction result for this event.
The transverse pressure $P_\mathrm{t}$
is a function of the vector potential ${A(x,y)}$ alone
and expressed by $P_\mathrm{t} = B^2_\mathrm{z}/2\mu + p$,
where $B_\mathrm{z}$ is the axial magnetic field
and $p$ is the thermal pressure
\citep{hs2002}.
The measurements are fitted by a cubic polynomial,
and an exponential tail is used for $A$ less than $-$220.8 T m.
The fitting residue $R_\mathrm{f}\approx0.04$ is relatively small,
which proves that the single-valued relation between $P_\mathrm{t}$ and $A$
required by the GS technique is satisfied
\citep{hsn2004JGRA}.
The GS reconstruction is sensitive to the chosen boundaries,
which can help determine the MC interval at \wind{}.

\section{Summary and Discussions} \label{discuss}
We have performed a comprehensive analysis of the \thecme,
covering the solar source observations by \sdo,
the stereoscopic remote-sensing observations from \stereo,
the magnetic field signatures at \mes,
and the type II radio burst and \insitu{} characteristics observed by \wind.
A GS reconstruction is employed to understand
the ICME structure
and how the structure controls the geomagnetic activity.
These results are summarized and discussed below,
which illustrate the importance of multi-spacecraft remote-sensing
and \insitu{} observations for understanding the physical processes
of CME propagation and space weather forecasting.

\begin{enumerate}
  \item This study compares
  \stereo{} stereoscopic wide-angle imaging observations
  with multi-point \insitu{} measurements at Mercury and the Earth,
  which places a strong constraint on CME Sun-to-Earth propagation.
  The CME kinematics determined from the triangulation technique
  predicts well the shock arrival time at \mes{}
  with an error of about 1 hour in this study.
  A reasonable accuracy is also obtained
  when we compare the predicted arrival time and speed
  with the \insitu{} measurements near the Earth.
  From the Sun to Earth, the CME undergoes an impulsive acceleration,
  a rapid deceleration, and then a gradual deceleration out to 1 AU,
  which agrees with the conclusions of
  \citet{lll2013,lhl2016}.
  We find that the rapid deceleration ceases at $\sim$50 R$_\sun$
  before the CME reached \mes{} ($\sim$0.47 AU),
  which is different from the coronagraph findings in
  \citet{gly2001}.
  Combining this case with the three events in
  \citet{lll2013},
  we suggest that fast CMEs are likely to terminate deceleration
  before reaching 100 R$_\sun$ from the Sun,
  which should be noticed in kinematics models of fast CMEs.
  This work once again proves the reliability
  of the stereoscopic triangulation technique
  in determining the kinematics of earthward CMEs in the inner heliosphere.

  \item Our comparison between wide-angle imaging observations
  and the interplanetary type II radio burst indicates important clues
  on the source region of the type II burst.
  The study reveals that the consistency between the shock distance
  derived from the type II radio burst
  and the CME kinematics determined by the triangulation method
  requires an unusually high solar wind electron density.
  The type II radio burst was probably
  produced from a high-density region.
  This discrepancy and the slow increasing distance
  of the CME-driven shock derived from the type II radio burst
  can be explained by the shock interacting with nearby streamers
  \citep[e.g.,][]{rvc2003ApJ,clm2007,fck2012ApJ}.
  Another possibility is that the type II radio burst
  was generated by the shock flank region with lower heights
  \citep{ca2002,kcr2003JGRA}.
  Because only one spacecraft detected the type II radio burst,
  it is not possible to determine the position of the source region
  by the radio triangulation method of
  \citet{mrb2015SoPh}.
  This result implies uncertainties in the determination
  of CME kinematics using type II radio bursts alone.

  \item We reconstruct the ICME structure near the Earth
  in order to connect it with the solar source observations
  and to understand how the structure contributes to the geomagnetic storm.
  The flux-rope inclination angle and chirality at 1 AU
  are consistent with those inferred from the observations of the solar source.
  It is worth noting that, however,
  a flux rope may rotate in the corona and interplanetary space
  \citep{ltl2010,vcn2011ApJ}.
  The prolonged southward magnetic field near the Earth
  is mainly from the axial component
  of the largely southward inclined flux rope.
  The axial magnetic field component of the flux rope
  is about two times as large as the azimuthal component
  as revealed by the GS reconstruction.
  If the flux rope had not been inclined southward
  to the large angle in this event,
  the strength and duration of the southward magnetic field
  would be much smaller.
  \citet{lhw2015} reported an event
  with the azimuthal magnetic field component
  much larger than the axial component,
  which suggests that a southward orientation is not a necessity
  for a strong southward magnetic field.
  These results indicate the importance
  of predicting both the flux-rope orientation
  and magnetic field components in geomagnetic activity forecasting.
\end{enumerate}

\acknowledgments
The research was supported by the Recruitment Program
of Global Experts of China, NSFC under grant 41374173
and the Specialized Research Fund for State Key Laboratories of China.
This study was supported by the Austrian Science Fund (FWF): [P26174-N27]
and the European Union Seventh Framework Programme (FP7/2007-2013)
under grant agreement No. 606692 [HELCATS].
The authors thank the referee and the editor
for their efforts to improve this article.
We acknowledge the use of data from {\stereo, \mes, \wind, \sdo},
and the D$_\mathrm{st}$ index from WDC in Kyoto.
% \facilities{}
% \software{}

% \notetoeditor{Figures 3 and 4 should appear in one page in print if possible.}
% \notetoeditor{Figures 9 and 10 should appear in one page in print if possible.}

\begin{figure}
\epsscale{.80}
\plotone{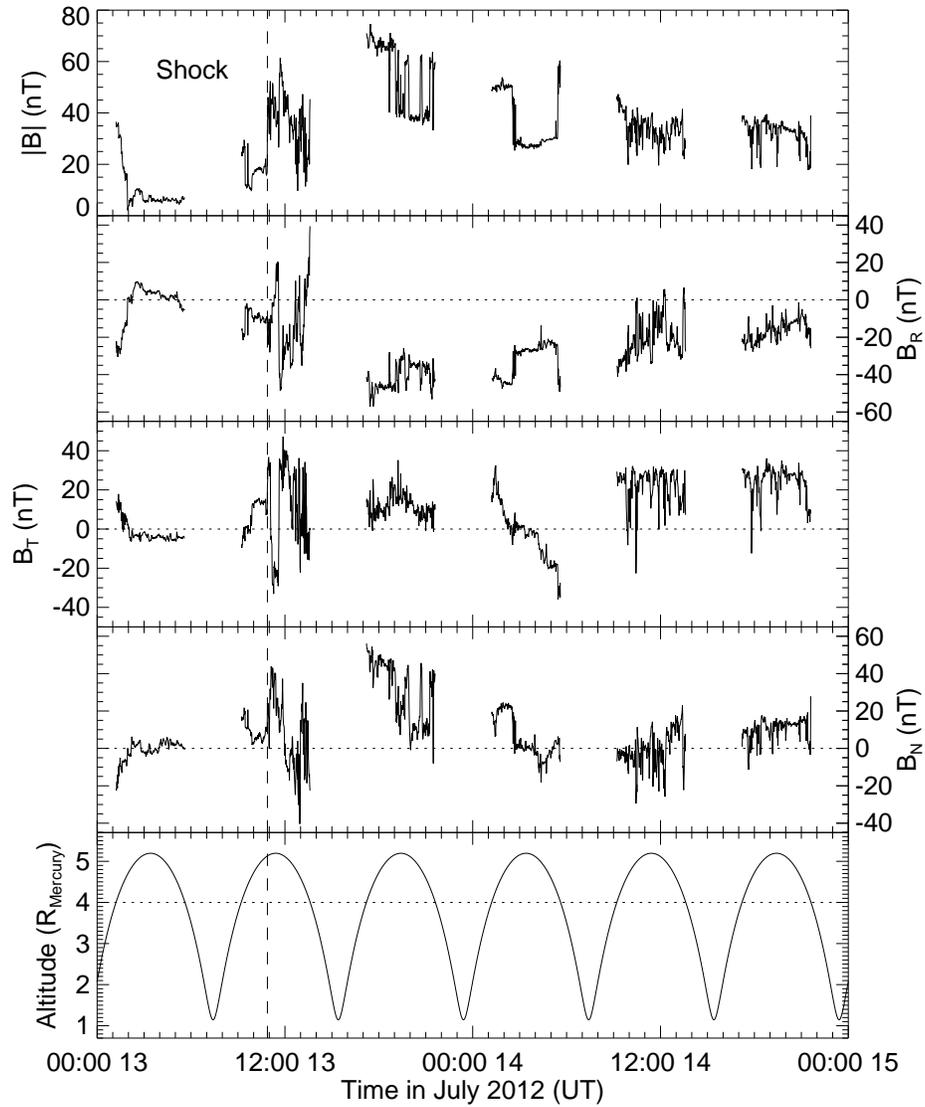}
\caption{\label{fig7}Magnetic field strength and components at \mes.
Also shown is the orbital altitude from the center of Mercury.
The vertical dashed line marks the arrival time
of the CME-driven shock at \mes.
The horizontal dotted line in the bottom panel
indicates the altitude (4 Mercury radii)
below which data are excluded.}
\end{figure}

\begin{figure}
\epsscale{.80}
\plotone{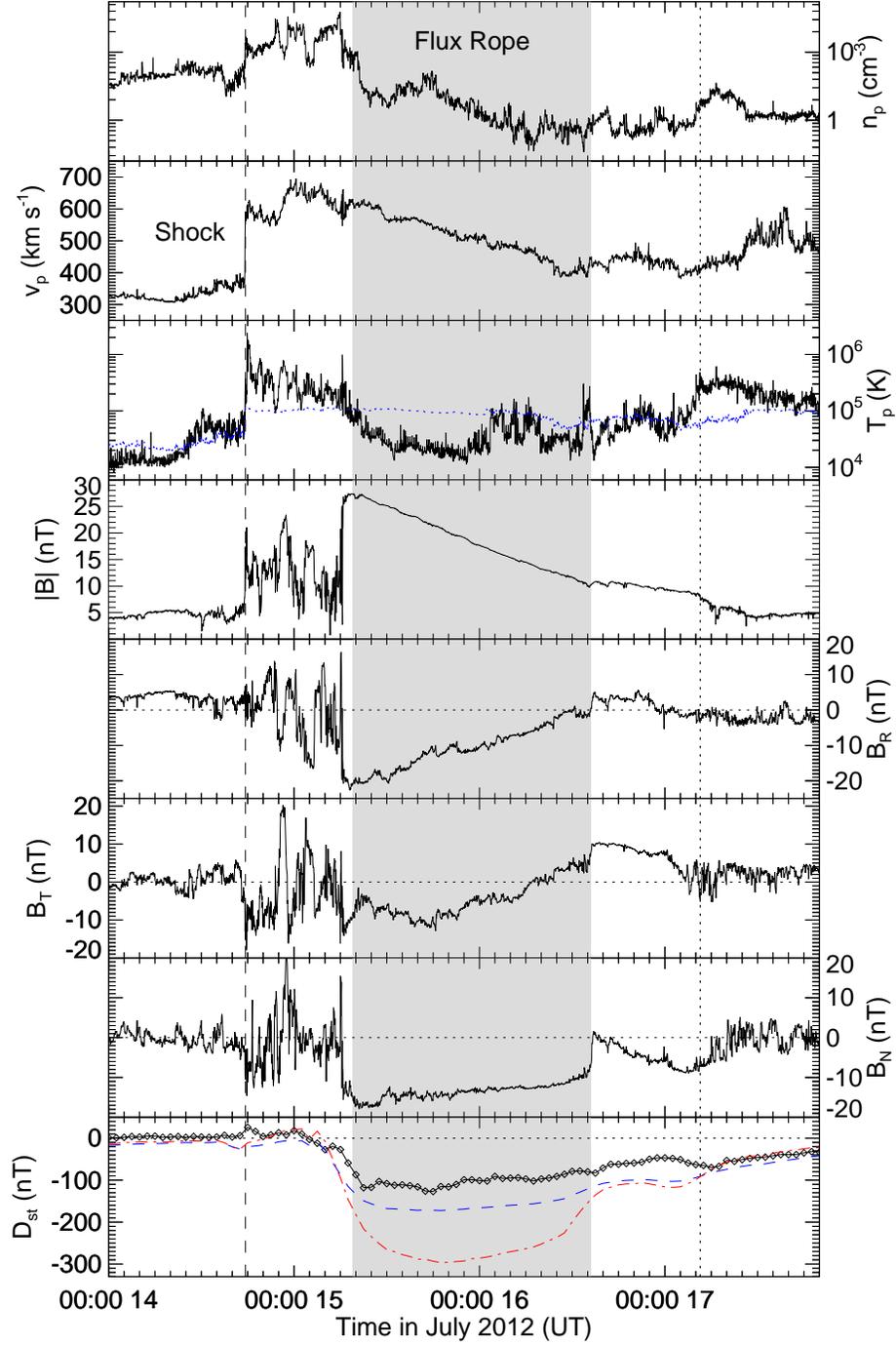}
\caption{\label{fig8}Solar wind plasma and magnetic field parameters
associated with the \thecme{} observed at \wind.
From top to bottom, the panels show the proton density,
bulk speed, proton temperature, magnetic field strength and components,
and D$_\mathrm{st}$, respectively.
The D$_\mathrm{st}$ profile is shifted 1 hour earlier
for comparing with the \insitu{} measurements.
The dotted line in the third panel
denotes the expected proton temperature calculated from the observed speed
\citep{lopez1987JGR}.
The shaded region indicates the magnetic cloud interval
determined by the GS reconstruction.
The vertical dashed and dotted lines
mark the arrival time of the shock
and the end of the magnetic cloud, respectively.
The blue dashed and red dot-dashed curves in the bottom panel
represent D$_\mathrm{st}$ values estimated
with the southward magnetic field component in GSM coordinates
using the formulae of
\citet{om2000JGR} and
\citet{bmr1975JGR}, respectively.}
\end{figure}

\begin{figure}
\epsscale{.80}
\plotone{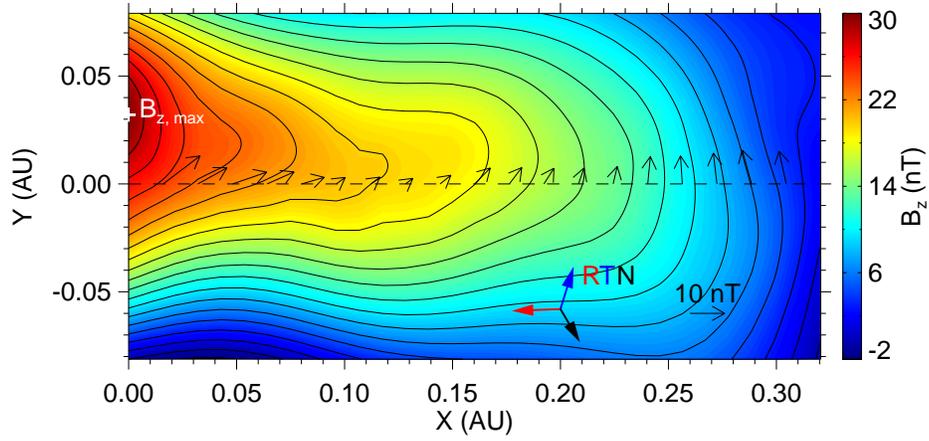}
\caption{\label{fig9}Reconstructed cross section
of the magnetic cloud at \wind.
Black contours show the distribution of the vector potential,
and the color shading indicates the value of the axial magnetic field
scaled by the color bar on the right.
The location of the maximum axial field is indicated by the white cross.
The dashed line marks the trajectory of \wind.
The thin black arrows denote the direction
and magnitude of the observed magnetic fields projected onto the cross section,
and the thick colored arrows show the projected RTN directions.}
\end{figure}

\begin{figure}
\epsscale{.80}
\plotone{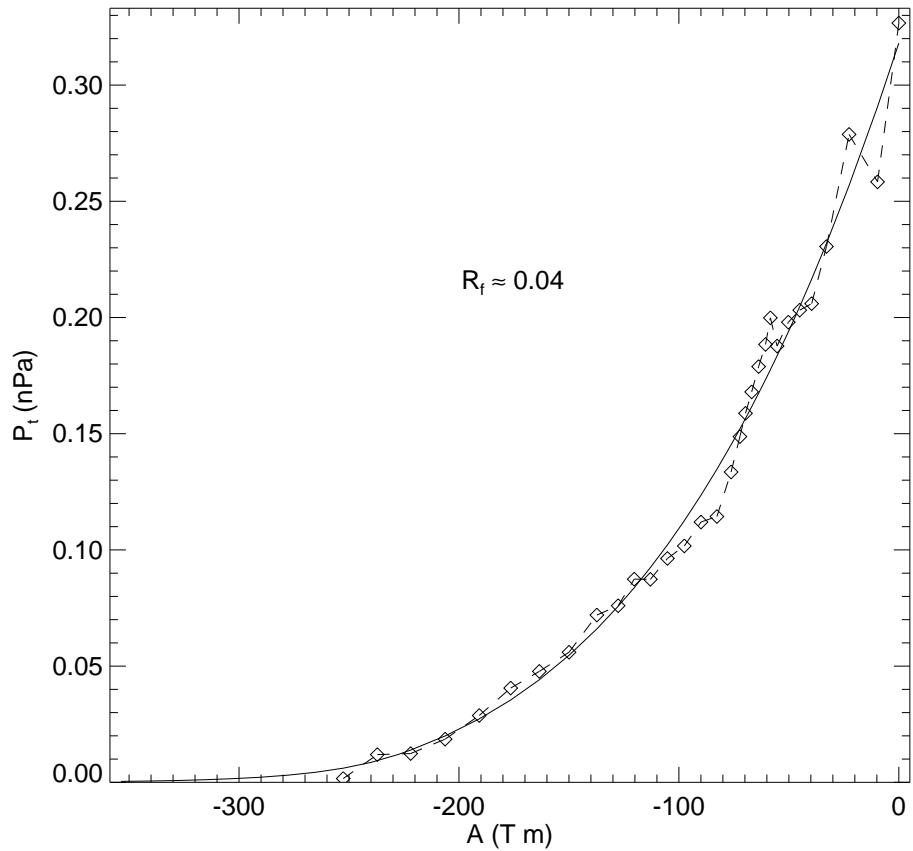}
\caption{\label{fig10}Transverse pressure $P_\mathrm{t}$
versus the vector potential $A$ along the trajectory of \wind.
The diamonds indicate measurements at \wind,
and the solid curve is a fitted cubic polynomial
combined with an exponential tail.
The fitting residue $R_\mathrm{f}$ is defined by \citet{hsn2004JGRA}.}
\end{figure}

\clearpage
\bibliography{hu.cme12jul12arxiv}
\bibliographystyle{aasjournal}
\end{document}